\documentclass{Interspeech2024}

\interspeechcameraready

\title{A Pilot Study of GSLM-based Simulation of Foreign Accentuation\\
Only Using Native Speech Corpora}

\name{Kentaro}{Onda}
\name{Joonyong}{Park}
\name{Nobuaki}{Minematsu}
\name{Daisuke}{Saito}

\address{
  Graduate School of Engineering, The University of Tokyo, Japan}
\email{\{ondakentaro,jpark,mine,dsk\_saito\}@gavo.t.u-tokyo.ac.jp}

\keywords{GSLM, foreign accentuation, self-supervised learning, language education,
World Englishes}

\usepackage{booktabs}
\usepackage{multirow}

\def\figref#1{\mbox{Figure \ref{#1}}}

\def\secref#1{\mbox{Section \ref{#1}}}

\let\pagerefsave=\pageref
\def\pageref#1{\mbox{\pagerefsave{#1}}}

\def\dqt#1{\mbox{\lq\lq}\protect\nolinebreak[4]{}#1\protect\nolinebreak[4]{}\mbox{\rq\rq}}

\begin{document}

\maketitle

\begin{abstract}
  %
  %
  We propose a method of simulating the human process of foreign accentuation
  using Generative Spoken Language Model (GSLM) only with native speech corpora.
  When one listens to spoken words of a foreign language and repeats them,
  the repeated speech is often with the accent of that listener's L1.
  This is said to be because the spoken words are mentally represented as a sequence of
  phonological units of the L1, and those units are used for oral reproduction.
  We simulate this process by inputting speech of language A
  into GSLM
  of language B to add B's accent onto the input speech.
  The process of running ASR of the L1 for foreign input speech
  and giving the ASR result to TTS of the L1 can be viewed as a naive implementation of this approach.
  The results of our experiments show that the synthesized accent
  of the output speech is highly natural,
  compared to real samples of A generated by speakers whose L1 is B,
  and that the degree
  of accentuation is
  controllable.

  %
\end{abstract}

\section{Introduction}
\label{sec:intro}


In recent years, the intelligibility principle has become
dominant in speech training for foreign language education, where learners do not always have to
acquire native-like pronunciation if their speech is intelligible
enough \cite{munro1995foreign,murphy2014intelligible,levis2020revisiting}.
This trend simply indicates that,
while the speaking side is allowed not to conform to a standardized pronunciation,
the listening side must adapt to diversely accented pronunciations. If we put a focus on
English education, research studies of World Englishes
estimate that about 74\% of English speakers are non-native \cite{kachru1992world,ethnologue},
and in international meetings such as INTERSPEECH
conferences, every attendee is surrounded by diversely accented Englishes.
How can we survive the diversity of accents in English?
To increase adaptability of listeners,
their exposure to World Englishes should be enhanced \cite{HVPT}. However, only a small number of
textbooks are available which provide ample speech samples with various accents \cite{nakanishi}.
One of the reasons of this situation is that recording educational passages read aloud with
various accents is costly, and this problem will be solved by introducing speech generation techniques which are flexible enough in controlling the kind and degree of foreign and/or regional
accents in the generated speech.

To the best of the authors' knowledge, however, techniques for converting text to accented speech
and those for converting native speech into accented speech generally require costly
accented non-native speech corpora \cite{liu2022controllable, bengali, quamer22_interspeech}.
However, considering how accented speech is generated by language learners,
this requirement seems a bit unnatural.
Learners generally learn English based only on native speech
without listening to, imitating, or training themselves to generate accented speech.
Non-native speech corpora are not needed for human learners to generate accented speech.
What is needed is just native corpora and learners' own mechanism of oral reproduction.
If we can simulate their own mechanism, non-native corpora will not be needed at all.



What is learners' own mechanism? In psycholinguistics, it is explained that
human listeners store in their mind an input acoustic stream of speech as sequence of phonological
units, which are often referred to as phonological representations \cite{flege1995second,Baddeley2007}.
If they have to reproduce the input speech orally, they convert the stored units back to acoustics
using their vocal organs. Due to this discretization and interpolation process, oral repetition
of speech input is totally different from that of non-speech input. While the extra-linguistic features in the input, such as gender and age, are not reproduced in the former, all the acoustic
aspects are replicated in the latter. The former is linguistic repetition, and the latter is
acoustic repetition, which does not undergo the discretization and interpolation process.
Now it is easy to understand why learners generate accented speech.
They represent an input foreign speech using phonological units of their L1, and when
they reproduce the input speech, their oral reproduction has to be with the L1's accent.


A naive way of simulating this process is running ASR of the L1 for foreign input speech and giving the ASR result to TTS of the L1. This approach lacks in its controllability in the
degree of accentuation, and in addition, we can claim a more essential deficit. Phonological representation
of speech is not phoneme representation or orthography of the speech, but the one with much more detailed information.
Although different researchers may claim differently on what should be contained in the phonological representation,
it seems to be a consensus that the phonological representation contains both segmental
and prosodic aspects of speech \cite{Goswami2012}. If good techniques are available to convert an input speech
as sequence of units which represent the both aspects, they will be used effectively
to simulate learners' own mechanism of oral repetition.

In this paper, we apply Generative Spoken Language Model (GSLM) to this purpose, which is
explained in detail in the following section. In GSLM,
the acquired units are derived from speech embeddings, which are considered to capture
the both aspects. Furthermore, the unique number of units is controllable.
Since GSLM-based resynthesis is a process of discretization and interpolation,
we input a speech sample of language A to GSLM of language B to add B's accent onto the input speech.
Considering some inevitable limitation of GSLM, as a pilot study, we examine and analyze
only the phonemic aspects of GSLM-based accented speech. Its quality and naturalness
are assessed by using real samples of accented speech.

\section{Generative Spoken Language Model}
\label{sec:GSLM}

\subsection{GSLM-based analysis-resynthesis}
\label{subsec:GSLM-resynthesis}

GSLM \cite{lakhotia-etal-2021-generative} is a model that replicates humans' ability to become
able to speak without explicit knowledge of any writing system, although they should acquire
some internal units for discretization and interpolation. Through self-supervised learning,
GSLM acquires what is called \dqt{units} and represents any input speech as a
sequence of adequately selected units.
%
%
\cite{lakhotia-etal-2021-generative} showed that the units can be used to construct a
unit-based language model,
sometimes referred to as \dqt{textless NLP}.
Furthermore, speech analysis-resynthesis was tested by concatenating Speech-to-unit (S2u) and unit-to-Speech (u2S). Experiments showed a comparable performance to concatenation of
existing ASR and TTS systems. In \cite{lakhotia-etal-2021-generative}, the experiments
were conducted within a language, but in this study, we apply the S2u (discretization) and u2S (interpolation) models of language B to speech samples of language A to simulate human accentuation.


\subsection{Speech-to-unit (S2u)}
\label{subsec:S2u}

S2u consists of two steps: 1) encoding of input speech with a representation learning model
and 2) discretization of the encoded speech by clustering.
Initially, windowed samples in an audio sequence are input to a pretrained representation learning model, 
which generates a vector for the samples.
Subsequently, the k-means clustering is performed to assign the vector to a specific cluster,
and the cluster index is treated as the unit.
In \cite{lakhotia-etal-2021-generative}, the best performance in the speech resynthesis task was obtained when HuBERT \cite{hubert} was used as the representation model,
and we use it in this study. For training the k-means clustering,
multiple speakers' samples are used.

\subsection{unit-to-Speech (u2S)}
\label{subsec:u2S}

A sequence of units is treated as text and an existing end-to-end TTS model is used for training and inference.
In \cite{lakhotia-etal-2021-generative}, Tacotron2 \cite{taco} was employed, and in this study,
we adopt the same model, and single-speaker speech corpora were used for training.

\section{GSLM-based foreign accentuation}
\label{sec:GSLM-accent}
\begin{figure}[t]
  \vspace*{-3mm}
  \centering

  \hspace*{-4mm}%
  \includegraphics[width=1.1\linewidth]{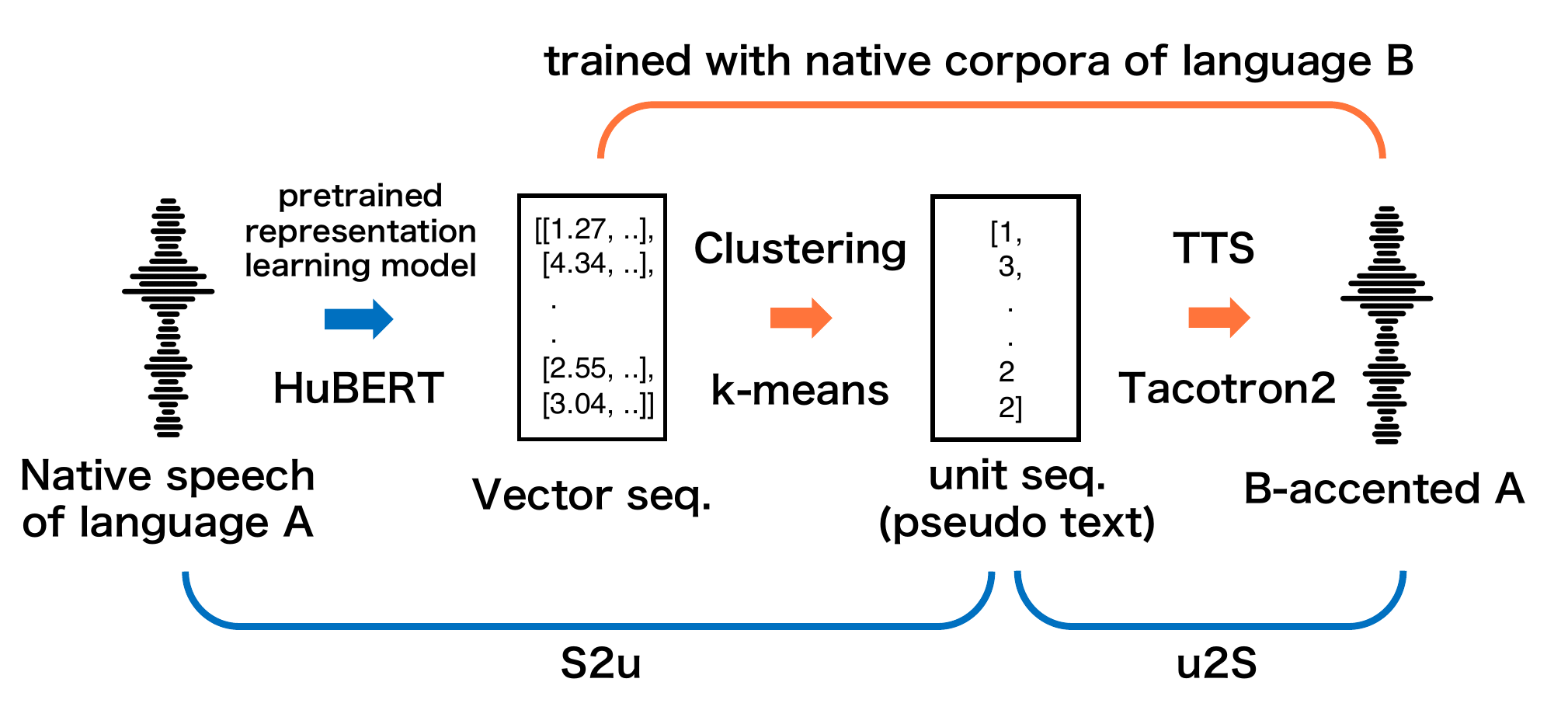}
  %
  %
  %

  \vspace*{-3mm}
  \caption{GSLM-based foreign accentuation}
  \label{fig:mechanism}
  \vspace*{-3mm}
\end{figure}
In this paper, as shown in \figref{fig:mechanism}, the pretrained HuBERT model is used to convert input speech of language A to a sequence of HuBERT vectors. The clustering model of S2u and the u2S model are, however, trained on language B.
It is expected that the u2S model generates
B-accented speech for input speech of language A\footnote{Sample audios are available at: \url{https://ondatk68.github.io/onda-demo/projects/accent_generation/} 
}, and for this
foreign accentuation, it is obvious that only native speech corpora are used.

%


We expect that concatenation of S2u and u2S will simulate well the human process of discretization and interpolation. If S2u$+$u2S are trained within language B to generate high-quality resynthesized speech but with as small units as possible,
the acquired set of units are regarded as a minimal set of units optimized for language B.
If a speech sample of language A is taken as input to our model of S2u(B)+u2S(B), since the input may be an outlier, S2u(B) will surely cause \dqt{linguistic discretization errors},
which are generally known as foreign accents. With this idea in mind, if the number of
units is increased, the degree of accentuation will be decreased. Although this expectation is examined
in the next section, within a language,
the number of units was shown to surely influence the quality of resynthesized speech
\cite{lakhotia-etal-2021-generative}.
Further, the linguistic discretization errors will depend on the systematic differences between the phonology of language A
and that of B.

Here we point out clear technical limitations of our model.
As explained in \secref{sec:intro}, the units acquired in our model are derived from
HuBERT embeddings, and they contain both the segmental and prosodic aspects of speech.
Our model, however, converts input to output frame by frame, where the temporal structure
of the input is always restored in the output. For this, our model cannot simulate
duration-based foreign accentuation at all. In the next section, we test our model and
analyze GSLM-based accented speech only in terms of the naturalness and quality of the
segmental, or phonemic aspect, where a special focus is put on phoneme substitutions
that are often discussed in applied linguistics \cite{Swan2001,You2005PronunciationVO,Kavanagh2007ThePO,MUNRO2006520}.

\section{Experiments}
\subsection{Experimental setup}

\begin{table}[tb]
  \vspace*{-3mm}
  \centering
  \caption{Corpora used for training our models}
  \label{tab:corpus}
  \vspace*{-3mm}

  \resizebox{\columnwidth}{!}{
    \begin{tabular}{llll}
      \toprule
      Language                 & S2u(multiple speakers)                                                                                   & u2S(single speaker)      \\\midrule 
      English                  & LibriSpeech \cite{libri}                                                                                 & LJSpeech \cite{ljspeech} \\        
      Japanese                 & JSUT \cite{sonobe2017jsut}, JVS \cite{takamichi2019jvs}, JKAC \cite{jkac}, JMAC \cite{takamichi2022jmac} & JSUT                     \\      
      Chinese                  & Primewords Chinese \cite{primewords201801}                                                               & BZNSYP \cite{bznsyp}     \\    
      Spanish                  & CSS10 \cite{park2019css10}, TEDx Spanish Corpus \cite{mena2019},                                         & CSS10   \\  
                               & West Point Heroico Spanish Speech \cite{heroico}                                                         &                          \\ 
      French                   & M-AILabs Speech Dataset \cite{mailab}                                                                    & CSS10                    \\ 
      \bottomrule
    \end{tabular}
  }
  \vspace*{-3mm}
\end{table}

Using the corpora shown in Table\ref{tab:corpus}, pairs of S2u and u2S were
constructed for American English, Japanese, Mandarin Chinese, Spanish, and French (EN, JPN, CHN, SPN, and FRC).
All the corpora consisted of native speech samples only,
and samples by multiple speakers were used for S2u and a single speaker for u2S.
The size of dataset used for each language was adjusted to be
about 100 hours for S2u and 10 hours for u2S by combining
multiple corpora or randomly selecting samples from a single corpus.
%
For each language, the number of units for S2u and u2S varied,
and it was 50, 200, or 1,000.
For training and inference of S2u, we used the implementation by Meta\footnote{\url{https://github.com/facebookresearch/fairseq/tree/main/examples/textless_nlp/gslm}},
and for u2S, we initially mapped each unit to a single Unicode character, and
then used this as input for Tacotron2 implemented by NVIDIA\footnote{\url{https://github.com/NVIDIA/tacotron2}}.
For all the languages,
the u2S models were trained in 100k steps with the default values for hyperparameters.
In the following experiments, we used American English samples as language A
and the above five languages as language B. If we use American English
for both A and B, it corresponds to GSLM-based analysis-resynthesis within a language,
whose performance will be used just as reference.


\begin{figure}[t]
  \centering
  \hspace*{-6mm}%
  \begin{minipage}{0.56\linewidth}
    \centering
    \includegraphics[width=\linewidth]{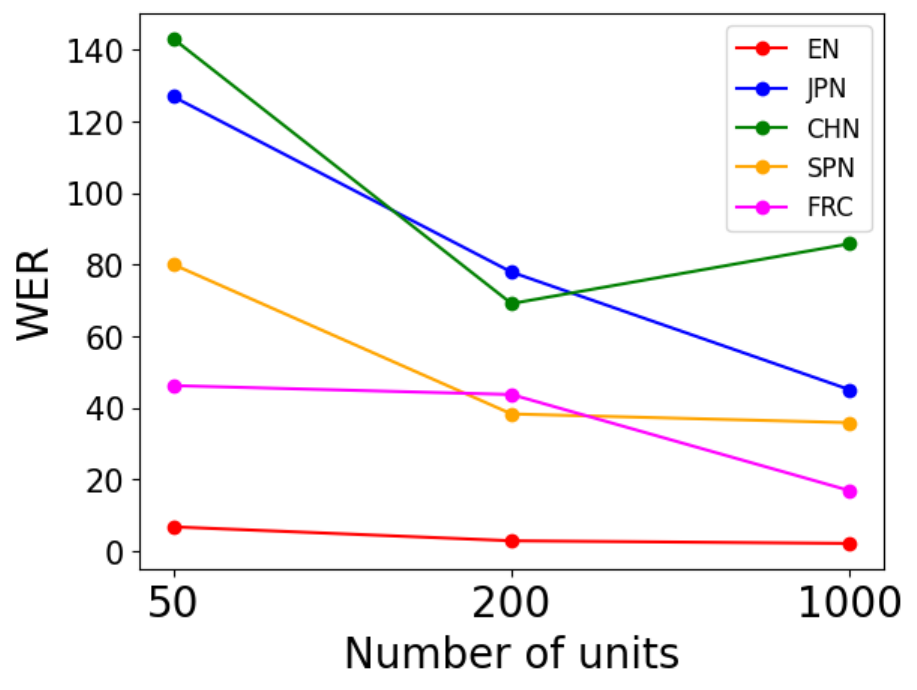}

    \vspace*{-2mm}
    \caption{Word Error Rate of the synthesized accented English}
    \label{fig:wer}
  \end{minipage}
  \hfill
  \begin{minipage}{0.56\linewidth}
    \centering
    \includegraphics[width=\linewidth]{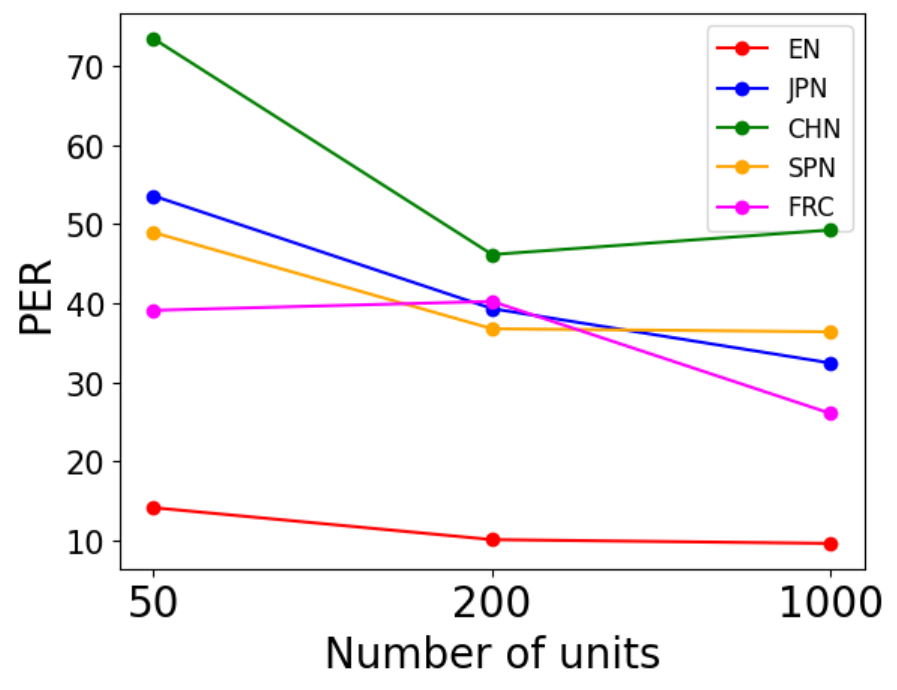}

    \vspace*{-2mm}
    \caption{Phone Error Rate of the synthesized accented English}
    \label{fig:per}
  \end{minipage}%
  \hspace*{-6mm}

  \vspace*{-5mm}
\end{figure}

\subsection{Degree of accent}
\label{subsec:degree_accent}

We ran the medium model of Whisper ASR\footnote{\url{https://github.com/openai/whisper}}
and the automatic phonetic transcriber of Allosaurus\footnote{\url{https://github.com/xinjli/allosaurus}} for all the input speech samples
and all the output speech obtained separately
by our five models of EN, JPN, CHN, SPN, and FRC.
As input speech, 5,934 audios of LJSpeech (American English), which were used for training the u2S model of EN,
were used.
By comparing the automatic transcription results of the input and
those of the output, Word Error Rate (WER) and Phone Error Rate (PER) were
calculated for each model. The former and the latter
quantify word-level intelligibility and phone-level deviation, respectively.
%
%
%
The results are shown in Figure \ref{fig:wer} and Figure \ref{fig:per}.
As expected, as the number of units increases,
both WER and PER tend to decrease, i.e., the degree of accent tends to decrease.
When comparing among languages, as for WER,
Spanish and French, which belong to the same Indo-European language family as English,
have lower accentuation than Japanese and Chinese, which are linguistically more distant.
As for PER, the interlingual differences are small,
it is especially interesting that Japanese scores are similar to Spanish and French.
This means that
each synthesized accent has equal phonetic
deviation from American English,
but in terms of word-level intelligibility,
European accents are easier to understand than Asian accents.

\subsection{Naturalness of accent in synthesized Japanese English}
\label{subsec:ERJ}

We evaluated the naturalness of the accents synthesized
in the output speech by comparing them with speech samples from
real non-native speakers using corpora where the same passages were
read aloud by American speakers and non-native speakers.
The American samples were used as input and the non-native ones
were compared with output, i.e. synthesized speech.
Before examining all the foreign accents, we first focused on Japanese English,
because a corpus is available, the individual speakers of which has
a large enough number of samples.



\subsubsection{English Read by Japanese (ERJ) corpus}

In the ERJ corpus \cite{Minematsu2004DevelopmentOE},
a shared set of sentences were read aloud by American speakers and Japanese learners.
A part of them, which are 460 phoneme-balanced sentences, were used here
and divided into 8 subsets so that all speakers in each subset read the same set of sentences.
%
%
In each subset, native English samples were used as input to within-language resynthesis,
and its output is denoted as synthesized American English (sAE).
The native samples were also resynthesized with the Japanese S2u$+$u2S models, which
is called synthesized Japanese English (sJE).
These were then compared phonetically with real American English
(rAE) and real Japanese English (rJE).

\subsubsection{Pronunciation deviation}

Pronunciation deviation of speaker $s$, PD$_s$, which characterizes the segmental
aspect of $s$'s accent,
was calculated objectively by quantifying each phoneme $p$'s pronunciation deviation from
the real native speakers in rAE.
\begin{align*}
   & \text{PD}_{s} = [\text{PD}_{s,1}, \text{PD}_{s,2}, \cdots, \text{PD}_{s,p}, \cdots, \text{PD}_{s,P}]                      \\
   & \text{PD}_{s,p} = \frac{\sum_{n \in \text{rAE}} \text{KL-divergence}(\text{APP}_{s,p} || \text{APP}_{n,p})}{|\text{rAE}|}
\end{align*}
$\text{PD}_{s,p}$ is the pronunciation deviation
observed when speaker $s$ intended phoneme $p$, and
$n$ is native speaker.
Here, $\text{APP}_{s,p}$ is the averaged distribution of phoneme-based posteriors
over the speech segments produced when speaker $s$ intended phoneme $p$, and
this is calculated for all $P$ phonemes.
The posteriors were calculated using WSJ-Kaldi \cite{Kaldi}.

\begin{figure}[tb]
  \centering
  \hspace*{-6mm}%
  \includegraphics[width=0.6\linewidth]{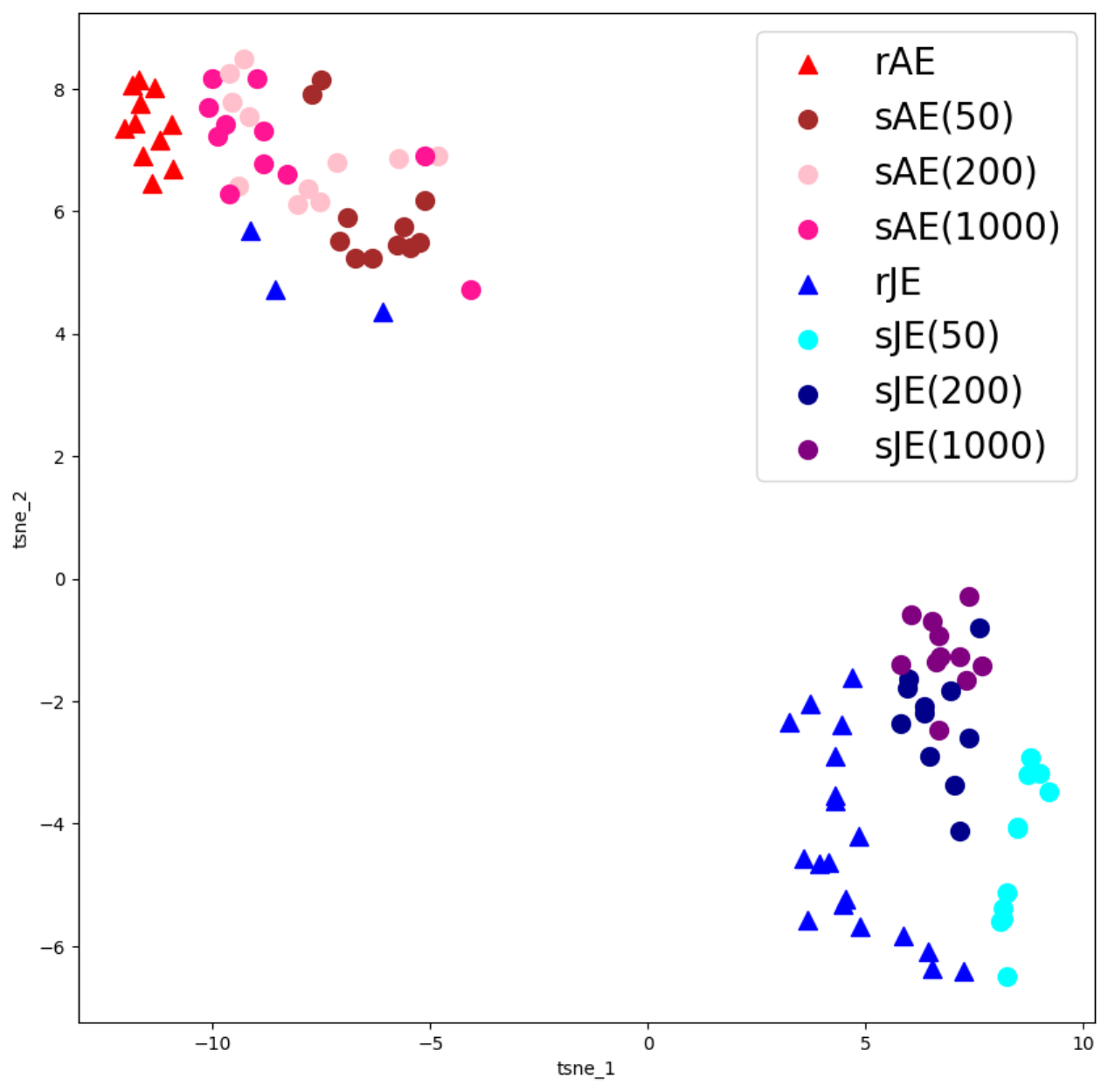}
  \vspace*{-3mm}
  \caption{Visualization of pronunciation deviations by t-SNE}
  \label{fig:tsne}
  \vspace*{-2mm}
\end{figure}

In Figure \ref{fig:tsne}, the distribution of \{PD$_s$\} of one of the subsets
is visualized using t-SNE \cite{tsne},
where each dot represents an output speaker or a real speaker.
This shows that sAE and sJE are close to rAE and rJE, respectively.
The synthesized speech has similar accent characteristics
to those of the real speakers.
Although not presented in this paper due to space limitation,
the other subsets showed very similar results.



Now, let us quantify the naturalness of accentuation in sJE by measuring its distance to
rJE in order to examine the effect of the number of units.
Here, the distance between a synthesized speaker $i$ ($i \in$sJE) and a real speaker
$j$ ($j \in$rJE) is calculated as correlation between PD$_i$ and PD$_j$. Using this, the
naturalness in accentuation of sJE, NA(sJE), is calculated as
\begin{equation*}
  \text{NA(sJE)} = \frac{\sum_{i\in \text{sJE}} \sum_{j\in \text{rJE}} \text{corr}(\text{PD}_i,\text{PD}_j)}{|\text{sJE}||\text{rJE}|}.
\end{equation*}
Table \ref{tab:corr} shows the averaged values of NA(sJE) over the subsets,
separately  for each number of units.
As reference, NA(rJE) is also shown, which is regarded as
the upper bound of NA(sJE).
%
%
The results show 
that NA(sJE) increases when the number of units increases, especially from 50 to 200.
It may seem contradictory that increasing the number of units reduces the degree of accent (as shown in \secref{subsec:degree_accent})
and also increases the naturalness of the accent at the same time.
This will be discussed in the next section.


\begin{table}[tb]
  \centering
  \caption{NA(rJE) and NA(sJE) for different numbers of units}
  \label{tab:corr}
  \vspace*{-2mm}
  \begin{tabular}{ccccc}
    \toprule
       & rJE  & sJE(50) & sJE(200) & sJE(1,000) \\\midrule
    NA & 0.77 & 0.57    & 0.67     & 0.67       \\
    \bottomrule
  \end{tabular}
  \vspace*{-3mm}
\end{table}

\subsection{Naturalness of accent in synthesized World Englishes}

To examine the naturalness of the accented English generated by the four (except English)
models,
we used a World Englishes speech corpus collected from speakers all over the world, who
read aloud the same one English passage.
Since the passage contains only 69 words, however,
to assess the synthesized accents,
we used a method different from the one used in the previous
section.
Here, we focused on phone substitution characteristics observed in real
accented English and synthesized accented English.

\subsubsection{Speech Accent Archive}
The Speech Accent Archive \cite{accentarchive} (SAA) is a dataset of the shared English
passage read aloud by a large number of native and non-native speakers of English.
This passage consists of 69 words and
is designed to include all the bi-phonemes typically found in General American,
making it suitable for comparing various accents.
\subsubsection{Phone substitution}
\label{subsec:SAA}
\begin{table}[tb]
  \centering
  \caption{Averaged phone substitution rate calculated from sJE}
  \label{tab:subrate}
  \vspace*{-2mm}


  \begin{tabular}{ccccc}  \toprule
    ranges of rJE       & count & sJE(50)        & sJE(200)       & sJE(1,000)     \\\midrule
    0.1$\leq$SR         & 15    & \textbf{0.152} & \textbf{0.156} & \textbf{0.155} \\
    0.05$\leq$SR$<$0.1  & 27    & 0.101          & \textbf{0.094} & \textbf{0.086} \\
    0.02$\leq$SR$<$0.05 & 65    & \textbf{0.049} & \textbf{0.045} & \textbf{0.042} \\
    0.01$\leq$SR$<$0.02 & 72    & 0.033          & 0.025          & 0.021          \\
    0$\leq$SR$<$0.01    & 829   & \textbf{0.010} & \textbf{0.008} & \textbf{0.006} \\  \bottomrule
  \end{tabular}
\end{table}

\begin{table}[t]
  \centering
  \caption{Ratio of substituting [s], [t], and [f] for [\texttheta]}
  \label{tab:theta}
  \vspace*{-3mm}
  \begin{tabular}{@{\,\,\,\,\,}c@{\,\,\,\,\,}c@{\,\,\,\,\,}c@{\,\,\,\,\,}c@{\,\,\,\,\,}c@{\,\,\,\,\,}c@{\,\,\,\,\,}}\hline
    Accent                    &     & real          & synth(50)     & synth(200)    & synth(1000)   \\\hline
    \multirow{3}{*}{Japanese} & [s] & \textbf{0.12} & 0.07          & 0.07          & 0.08          \\
                              & [t] & 0.06          & \textbf{0.24} & \textbf{0.22} & \textbf{0.15} \\
                              & [f] & 0.01          & 0.02          & 0.04          & 0.04          \\\hline
    \multirow{3}{*}{Chinese}  & [s] & \textbf{0.09} & \textbf{0.12} & \textbf{0.15} & \textbf{0.13} \\
                              & [t] & 0.04          & 0.05          & 0.08          & 0.06          \\
                              & [f] & 0.03          & 0.04          & 0.05          & 0.04          \\\hline
    \multirow{3}{*}{Spanish}  & [s] & 0.03          & 0.04          & 0.05          & 0.06          \\
                              & [t] & \textbf{0.10} & \textbf{0.29} & \textbf{0.24} & \textbf{0.27} \\
                              & [f] & 0.03          & 0.05          & 0.08          & 0.05          \\\hline
    \multirow{3}{*}{French}   & [s] & 0.04          & 0.02          & 0.02          & 0.02          \\
                              & [t] & \textbf{0.11} & \textbf{0.19} & \textbf{0.20} & \textbf{0.28} \\
                              & [f] & 0.06          & 0.08          & 0.10          & 0.11          \\\hline
  \end{tabular}
  \vspace*{-3mm}
\end{table}
Initially, all the 431 oral passages by American speakers, rAE, were accentuated
by the proposed S2u$+$u2S method to generate sJE, sCE, sSE, and sFE.
Then, all the input and output utterances were phonetically transcribed
by Allosaurus \cite{allosaurus}.
Subsequently, the speech sample named as english81 was selected as model
speech because the speaker is from
the Midwestern region of US, where General American is spoken.
By comparing each of the non-native samples with the model speech,
phone-based substitution rate, SR, was defined for each pair of original phone in the model speech, $p_o$, and substituted phone in s$X$E or r$X$E, $p_s$, where $p_s$$\not =$$p_o$ and $X$$=$J, C, S or F, as
    \vspace*{-0.2mm}
    \footnotesize
    \begin{equation*}
      \text{SR}(p_s | p_o; Y) = \frac{\text{Number of } p_s \text{ substituted for } p_o \text{in Y}}
      {\text{Number of } p_o \text{ in the model speech}},
    \end{equation*}
    \normalsize
    where $Y$ is $x$JE, $x$CE, $x$SE, or $x$FE. $x$ is synthesized or real.
    Next, for all approximately 1,000 phone pairs of $p_s$ and $p_o$, $\text{SR}(p_s | p_o; \text{r$X$E})$ and
  $\text{SR}(p_s | p_o; \text{s$X$E})$ were calculated for language $X$.
    Then, the phone pairs were grouped based on
  $\text{SR}(p_s | p_o; \text{r$X$E})$
    into the following five ranges of
    1) [0.1, $-$) , 2) [0.05, 0.1), 3) [0.02, 0.05), 4) [0.01, 0.02), 5) [0, 0.01).
    After that, for each range, its values of $\text{SR}(p_s | p_o; \text{s$X$E})$, not  $\text{SR}(p_s | p_o; \text{r$X$E})$, are
    averaged over all the phone pairs in that range.
    Table \ref{tab:subrate} shows the number of phone pairs and the averaged $\text{SR}(p_s | p_o; \text{s$X$E})$ for each range in the case of Japanese English.
    The higher the SR for rJE, the fewer the kinds of phone pairs and the higher the SR for sJE.
    %
    In almost all cases, the SR averages of sJE fall in the same ranges as rJE (shown in bold), and
    even they are not, the differences are very small.
    We can say that sJE's phone substitution characteristics are close to rJE.
    Although Table \ref{tab:subrate} shows the results of Japanese English, similar results
    were obtained with other three accented Englishes.

    Focusing on the effect of the number of units, when 0.1$\leq$SR, the average SR of sJE does not
    decrease even when the number of units increases, unlike the other ranges.
    As shown in Table \ref{tab:subrate}, however, this range has only a small number of substitutions,
    so the overall degree of accentuation decreases as the number of units increases, as shown in \secref{subsec:degree_accent}.
    These frequent, and typical substitutions are thought to be caused by fatal differences in the phonological systems
    between L1 and L2, and thus the ``linguistic discretization errors'' cannot be interpolated only by increasing the number of units.
    Other substitutions may be based on linguistic differences that are less significant or
    some speaker-specific differences that are not shared
    among real L2 speakers.
    These types of errors were reduced by increasing the number of units.
    Here, the question posed at the end of \secref{subsec:ERJ} is answered
    by explaining that
    typical and characteristic phone substitutions, which happen frequently,
    still remain even with a higher number of units, although
    other ``noisy'' substitutions are reduced by increasing the number of units,
    resulting in more natural, ``pure" accentuation.

    %
    %
    Lastly, we take [\texttheta] as an example to show that our method does indeed reproduce
    accentuation tendencies that are actually found in real L2 speakers. \cite{th} explains
    that [\texttheta] is
    often replaced by [s], [t], or [f] in L2 speakers' utterances.
    %
    Table \ref{tab:theta} shows $\text{SR}(p_s | \text{\texttheta})$ for [s], [t], and [f]
    in both cases of real L2 speech and synthesized L2 speech.
    %
    The most frequently substituted phones are shown in bold.
    Looking at real speakers' results of Table \ref{tab:theta},
  $\text{SR}(\text{s} | \text{\texttheta})$$>$$\text{SR}(\text{t} | \text{\texttheta})$$>$$\text{SR}(\text{f} | \text{\texttheta})$ for Japanese and Chinese,
  $\text{SR}(\text{t} | \text{\texttheta})$$>$$\text{SR}(\text{s} | \text{\texttheta})$$\fallingdotseq$$\text{SR}(\text{f} | \text{\texttheta})$ for Spanish, and
  $\text{SR}(\text{t} | \text{\texttheta})$$>$$\text{SR}(\text{f} | \text{\texttheta})$$>$$\text{SR}(\text{s} | \text{\texttheta})$ for French.
For all languages except for Japanese, the proposed method successfully reproduces
the real substitution tendencies in synthesized accented speech.
This suggests that the method has learned and replicated L1-dependent substitution tendencies
of [\texttheta].
As mentioned above, the increased number of units does not affect the SRs for [\texttheta],
showing that the substitutions of [\texttheta] characterize each accent.




\section{Conclusion}
In this paper,
a novel method was introduced for the first time to simulate foreign accentuation
using GLSM and solely native speech corpora.
As a pilot study, we have shown the validity of GSLM for synthesizing accented speech
through the analysis from a phonemic perspective.
The synthesized accents realized well
the characteristics found in real accents
and interestingly, as the number of units is increased, the overall degree of accent decreases,
but the typical phone substitutions are still found in the synthesized accent,
resulting in a more natural replication of the accent.
This is the first step toward more natural accented speech synthesis, and further work is needed
to replicate the duration-based accents and to analyze the output speech from other perspectives
than phonemic ones.





\bibliographystyle{IEEEtran}
\bibliography{references}

\begin{thebibliography}{10}
\providecommand{\url}[1]{#1}
\csname url@samestyle\endcsname
\providecommand{\newblock}{\relax}
\providecommand{\bibinfo}[2]{#2}
\providecommand{\BIBentrySTDinterwordspacing}{\spaceskip=0pt\relax}
\providecommand{\BIBentryALTinterwordstretchfactor}{4}
\providecommand{\BIBentryALTinterwordspacing}{\spaceskip=\fontdimen2\font plus
\BIBentryALTinterwordstretchfactor\fontdimen3\font minus \fontdimen4\font\relax}
\providecommand{\BIBforeignlanguage}[2]{{%
\expandafter\ifx\csname l@#1\endcsname\relax
\typeout{** WARNING: IEEEtran.bst: No hyphenation pattern has been}%
\typeout{** loaded for the language `#1'. Using the pattern for}%
\typeout{** the default language instead.}%
\else
\language=\csname l@#1\endcsname
\fi
#2}}
\providecommand{\BIBdecl}{\relax}
\BIBdecl

\bibitem{munro1995foreign}
M.~J. Munro and T.~M. Derwing, ``Foreign accent, comprehensibility, and intelligibility in the speech of second language learners,'' \emph{Language Learning}, vol.~45, no.~1, pp. 73--97, 1995.

\bibitem{murphy2014intelligible}
J.~M. Murphy, ``Intelligible, comprehensible, non-native models in esl/efl pronunciation teaching,'' \emph{System}, vol.~42, pp. 258--269, 2014.

\bibitem{levis2020revisiting}
J.~Levis, ``Revisiting the intelligibility and nativeness principles,'' \emph{Journal of Second Language Pronunciation}, vol.~6, no.~3, pp. 310--328, 2020.

\bibitem{kachru1992world}
B.~B. Kachru, ``World englishes: approaches, issues and resources,'' \emph{Language Teaching}, vol.~25, no.~1, pp. 1--14, 1992.

\bibitem{ethnologue}
``What is the most spoken language? | ethnologue,'' \url{https://www.ethnologue.com/insights/most-spoken-language/}.

\bibitem{HVPT}
H.~Zhang, Y.~Inoue, D.~Saito, N.~Minematsu, and Y.~Yamauchi, ``Computer-aided high variability phonetic training to improve robustness of learners’listening comprehension,'' in \emph{Proceedings of the 19th International Congress of Phonetic Sciences (ICPhS)}, 2019.

\bibitem{nakanishi}
N.~Nakanishi, ``Sounds of englishes: English pronunciation database,'' \emph{Acoustical Society of Japan 2020 Fall Meeting}, 2020.

\bibitem{liu2022controllable}
R.~Liu, B.~Sisman, G.~Gao, and H.~Li, ``Controllable accented text-to-speech synthesis,'' 2022.

\bibitem{bengali}
S.~Chandra, P.~Bharati, and S.~K.~D. Mandal, ``Towards the development of accent conversion model for (l1)bengali speaker using cycle consistent adversarial network (cyclegan),'' in \emph{2022 25th Conference of the Oriental COCOSDA International Committee for the Co-ordination and Standardisation of Speech Databases and Assessment Techniques (O-COCOSDA)}, 2022, pp. 1--5.

\bibitem{quamer22_interspeech}
W.~Quamer, A.~Das, J.~Levis, E.~Chukharev-Hudilainen, and R.~Gutierrez-Osuna, ``{Zero-Shot Foreign Accent Conversion without a Native Reference},'' in \emph{Proc. Interspeech 2022}, 2022, pp. 4920--4924.

\bibitem{flege1995second}
J.~E. Flege, ``Second language speech learning: Theory, findings, and problems,'' \emph{Speech perception and linguistic experience: Issues in cross-language research}, vol.~92, pp. 233--277, 1995.

\bibitem{Baddeley2007}
A.~Baddeley, \emph{{Working memory, thought, and action}}.\hskip 1em plus 0.5em minus 0.4em\relax Oxford University Press, 2007.

\bibitem{Goswami2012}
U.~Goswami, ``{Phonological Representation},'' in \emph{Encyclopedia of the Sciences of Learning}, N.~M. Seel, Ed.\hskip 1em plus 0.5em minus 0.4em\relax Boston, MA: Springer US, 2012, pp. 2625--2627.

\bibitem{lakhotia-etal-2021-generative}
\BIBentryALTinterwordspacing
K.~Lakhotia, E.~Kharitonov, W.-N. Hsu, Y.~Adi, A.~Polyak, B.~Bolte, T.-A. Nguyen, J.~Copet, A.~Baevski, A.~Mohamed, and E.~Dupoux, ``On generative spoken language modeling from raw audio,'' \emph{Transactions of the Association for Computational Linguistics}, vol.~9, pp. 1336--1354, 2021. [Online]. Available: \url{https://aclanthology.org/2021.tacl-1.79}
\BIBentrySTDinterwordspacing

\bibitem{hubert}
W.-N. Hsu, B.~Bolte, Y.-H.~H. Tsai, K.~Lakhotia, R.~Salakhutdinov, and A.~Mohamed, ``Hubert: Self-supervised speech representation learning by masked prediction of hidden units,'' \emph{IEEE/ACM Transactions on Audio, Speech, and Language Processing}, vol.~29, pp. 3451--3460, 2021.

\bibitem{taco}
J.~Shen, R.~Pang, R.~J. Weiss, M.~Schuster, N.~Jaitly, Z.~Yang, Z.~Chen, Y.~Zhang, Y.~Wang, R.~Skerrv-Ryan, R.~A. Saurous, Y.~Agiomvrgiannakis, and Y.~Wu, ``Natural tts synthesis by conditioning wavenet on mel spectrogram predictions,'' in \emph{2018 IEEE International Conference on Acoustics, Speech and Signal Processing (ICASSP)}, 2018, pp. 4779--4783.

\bibitem{Swan2001}
M.~Swan and B.~Smith, \emph{Learner English --A teacher's guide to interference and other problems--}.\hskip 1em plus 0.5em minus 0.4em\relax Cambridge University Press, 2001.

\bibitem{You2005PronunciationVO}
\BIBentryALTinterwordspacing
H.~You, A.~Alwan, E.~A. Kazemzadeh, and S.~S. Narayanan, ``Pronunciation variations of spanish-accented english spoken by young children,'' in \emph{Interspeech}, 2005. [Online]. Available: \url{https://api.semanticscholar.org/CorpusID:16037316}
\BIBentrySTDinterwordspacing

\bibitem{Kavanagh2007ThePO}
\BIBentryALTinterwordspacing
B.~Kavanagh, ``The phonemes of japanese and english : A contrastive analysis study,'' 2007. [Online]. Available: \url{https://api.semanticscholar.org/CorpusID:59362852}
\BIBentrySTDinterwordspacing

\bibitem{MUNRO2006520}
\BIBentryALTinterwordspacing
M.~J. Munro and T.~M. Derwing, ``The functional load principle in esl pronunciation instruction: An exploratory study,'' \emph{System}, vol.~34, no.~4, pp. 520--531, 2006. [Online]. Available: \url{https://www.sciencedirect.com/science/article/pii/S0346251X06000856}
\BIBentrySTDinterwordspacing

\bibitem{libri}
V.~Panayotov, G.~Chen, D.~Povey, and S.~Khudanpur, ``Librispeech: An asr corpus based on public domain audio books,'' in \emph{2015 IEEE International Conference on Acoustics, Speech and Signal Processing (ICASSP)}, 2015, pp. 5206--5210.

\bibitem{ljspeech}
K.~Ito and L.~Johnson, ``The lj speech dataset,'' \url{https://keithito.com/LJ-Speech-Dataset/}, 2017.

\bibitem{sonobe2017jsut}
R.~Sonobe, S.~Takamichi, and H.~Saruwatari, ``Jsut corpus: free large-scale japanese speech corpus for end-to-end speech synthesis,'' 2017.

\bibitem{takamichi2019jvs}
S.~Takamichi, K.~Mitsui, Y.~Saito, T.~Koriyama, N.~Tanji, and H.~Saruwatari, ``Jvs corpus: free japanese multi-speaker voice corpus,'' 2019.

\bibitem{jkac}
S.~Takamichi, ``Japanese kamishibai and audiobook corpus (j-kac),'' 2021.

\bibitem{takamichi2022jmac}
S.~Takamichi, W.~Nakata, N.~Tanji, and H.~Saruwatari, ``J-mac: Japanese multi-speaker audiobook corpus for speech synthesis,'' 2022.

\bibitem{primewords201801}
L.~Primewords Information Technology~Co., ``Primewords chinese corpus set 1,'' 2018, \url{https://www.primewords.cn}.

\bibitem{bznsyp}
L.~Databaker (Beijing) Technology~Co., ``Chinese standard mandarin speech copus,'' 2021, \url{https://www.data-baker.com/open_source.html}.

\bibitem{park2019css10}
K.~Park and T.~Mulc, ``Css10: A collection of single speaker speech datasets for 10 languages,'' 2019.

\bibitem{mena2019}
C.~D. Hernandez-Mena, ``Tedx spanish corpus: Audio and transcripts in spanish taken from the tedx talks; shared under the cc by-nc-nd 4.0 license,'' Web Download, 2019.

\bibitem{heroico}
L.~D.~C. John~Morgan, ``West point heroico spanish speech,'' 2006.

\bibitem{mailab}
\BIBentryALTinterwordspacing
I.~S. M-AILABS, ``M-ailabs speech dataset,'' 2018. [Online]. Available: \url{https://www.caito.de/2019/01/03/the-m-ailabs-speech-dataset/}
\BIBentrySTDinterwordspacing

\bibitem{Minematsu2004DevelopmentOE}
\BIBentryALTinterwordspacing
N.~Minematsu, Y.~Tomiyama, K.~Yoshimoto, K.~Shimizu, S.~Nakagawa, M.~Dantsuji, and S.~Makino, ``Development of english speech database read by japanese to support call research,'' 2004. [Online]. Available: \url{https://api.semanticscholar.org/CorpusID:6642302}
\BIBentrySTDinterwordspacing

\bibitem{Kaldi}
D.~Povey, A.~Ghoshal, G.~Boulianne, L.~Burget, O.~Glembek, N.~Goel, M.~Hannemann, P.~Motl^^c3^^ad^^c4^^8dek, Y.~Qian, P.~Schwarz, J.~Silovsk^^c3^^bd, G.~Stemmer, and K.~Vesel, ``The kaldi speech recognition toolkit,'' \emph{IEEE 2011 Workshop on Automatic Speech Recognition and Understanding}, 01 2011.

\bibitem{tsne}
L.~Van~der Maaten and G.~Hinton, ``Visualizing data using t-sne.'' \emph{Journal of machine learning research}, vol.~9, no.~11, 2008.

\bibitem{accentarchive}
\BIBentryALTinterwordspacing
S.~Weinberger, ``Speech accent archive,'' 2015. [Online]. Available: \url{http://accent.gmu.edu}
\BIBentrySTDinterwordspacing

\bibitem{allosaurus}
X.~Li, S.~Dalmia, J.~Li, M.~Lee, P.~Littell, J.~Yao, A.~Anastasopoulos, D.~R. Mortensen, G.~Neubig, A.~W. Black, and F.~Metze, ``Universal phone recognition with a multilingual allophone system,'' in \emph{ICASSP 2020 - 2020 IEEE International Conference on Acoustics, Speech and Signal Processing (ICASSP)}, 2020, pp. 8249--8253.

\bibitem{th}
E.~Koffi, ``The pronunciation of voiceless th in seven varieties of l2 englishes: Focus on intelligibility,'' \emph{Linguistic Portfolios}, vol.~4, 2015.

\end{thebibliography}

\end{document}